# THE PROCESS AND PATHOGEN BEHAVIOUR IN COMPOSTING: A REVIEW

N. M.SUNAR, [1, 2, *], STENTIFORD, E.I.[1], STEWART, D.I [1], AND FLETCHER, L.A.[1]

[1] Pathogen Control Engineering Research Group, School Civil Engineering, University of Leeds, Leeds, LS2 9JT, United Kingdom
[2] Unit of Bioengineering, Faculty of Civil and Environmental Engineering, University Tun Hussein Onn Malaysia, 86400 Batu Pahat, Johor, Malaysia
[*] Corresponding Author. Tel: +44 (0) 113 3432319. Email: shuhaila@uthm.edu.my

**Abstract**

*Composting is defined as the biological decomposition and stabilization of organic substrates under aerobic conditions to allow the development of thermophilic temperatures. This thermophilic temperature is a result of biologically produced heat. Composting produces the final product which is sufficiently stable for storage and application to land without adverse environmental effects. There are many factors which affect the decomposition of organic matter in the composting process. Since the composting process is very intricate, it is not easy to estimate the effect of a single factor on the rate of organic matter decomposition. This paper looked at the main factors affecting the composting process. Problems regarding the controlling, inactivation and regrowth of pathogen in compost material are also discussed.*

## 1 Introduction

Composting referred to as an aerobic, biological process that uses naturally occurring microorganisms to convert mixture of organic matter into a humus-like product (van Elsas et al., 2007). This process destroys pathogens, converts nitrogen from unstable ammonia to stable organic forms, and reduces the volume of waste (Imbeah, 1998). According to its etymological meaning, *composting* comes from Latin *compositum* which is meaning mixture. This word refers to biodegradation process of mixture of substrates in aerobic conditions and solid state. Compost is the main product of composting, which is defined as the stabilized, sanitized, compatible and beneficial to plant growth. According to Bertoldi (2007), compost has undergone:

     a- An initial, rapid stage of decomposition
     b- A stage of stabilization
     c- An incomplete process of humification.

Meanwhile, described by Diaz et al., (2007) the essential need is to transform fresh organic matter into compost mainly for three reasons:

     a- To overcome the phytotoxicity of fresh non-stabilized organic matter;
     b- To reduce the presence of agents (viruses, bacteria, fungi, parasites ) that are pathogenic to man, animals and plants to a level that does not further constitute a health risk; and
     c- To produce an organic fertilizer or a soil conditioner, recycling organic wastes and biomass.

## 2    Factors which influence the composting process

Breakdown of material composition in composting occurs as a natural process. But, the length of time and modalities for natural composting will require a long and heterogeneous process, making it unsuitable for industrial use. According to Ahn et al., (2008) enumerated more than 20 factors affect the decomposition of organic matter in the composting process. As a result of this, the composting process is very intricate, so it's not easy to estimate the effect of a single factor on the rate of organic decomposition. However, several factors were generally recognized as primary factors affecting the composting process. These factors contribute to an optimum environment for the microbial process in composting, which can be made into functions of technology as follows:

### 2.1    Nutrient content

Material composition in term of nutrient content is an important strategy in controlling the composting process. The degradability of organic material in composting depends on the chemical composition of its organic matter fraction, as well as on the concentration of the C/N fraction. Normally, organic wastes differ from each other as regards their content of nitrogen. For example for manure (1.2–6.3 g N/100 g dry matter), food wastes (3.2 g N/100 g DM), grass clippings (2–6 g N/100 g DM), and wastewater sludge (2–7 g N/100 g DM) usually have high nitrogen contents, meanwhile cellulose materials such as sawdust (0.1 g N/100 g DM), straw (0.3–1 g N/100 g DM), and leaves (0.5–1 g N/100 g DM) are quite low in nitrogen (de Guardia et al., 2008).

The processes of controlling composting are made much easier by determine a suitable C/N value. Microorganisms, utilized for their metabolism have approximately 30 part of carbon for each part of nitrogen. Around 20 part of carbon are oxidized to $CO_2$ (ATP), and 10 parts are utilized in synthesising protoplasm (Diaz et al., 2007b). By assuming microorganisms use 15–30 parts of carbon for each part of nitrogen, an optimum C/N ratio ranging between 15 and 30 was recommended for rapid composting (Huang et al., 2004, de Guardia et al., 2008). The C/N has an important effect on the composting process. The higher C/N values would also have the effect of limiting the growth of microorganisms, resulting in the composting process being slowed down. On the other hand, lower C/N would lead to excess available nitrogen and higher losses as ammonia in the exhaust gases (Sommer and Dahl, 1999, de Guardia et al., 2008, Pagans et al., 2006). This may lead to an alleviation of the odour problem that is usually encountered in full-scale composting facilities (Li et al., 2008).

## 2.2   Moisture content

Moisture content is the most important parameter in composting and as it greatly affects the rate of decomposition. Traditionally, composting systems have been controlled by on temperature, however, previous studies proved that moisture content effects are more influential than temperature (Liang et al., 2003). Stentiford (1996) stated that the moisture content of the compost affects (a) the structural properties of the materials (b) the thermal properties of the material and (c) the rate of biodegradation. Moisture in the composting process is effected by microbial activity and thus influences temperature and the rate of decomposition (Turner, 2002). Moisture content also provides a medium for the transport of dissolved nutrients required for the metabolic and physiological activities of microorganisms (Miller, 1989, Stentiford, 1996) This physical parameter is closely related to aeration rates in terms of displacement of air in the interstices by water and promotion of clumping as well as lowering of the structural strength of the material (de Bertoldi et al., 1983).

The optimization of moisture content in composting varies with the size of the particles and physical state of the material. Each material of composting has unique physical characteristics, and these affect the relationship between moisture content and its corollary factors, water availability, particle sizes, porosity, and permeability (Ahn et al., 2008).

Moisture content always serves as a proxy for other critical factors such as water availability and microbial activity in different moisture ranges. The very low moisture content values would cause early dehydration during composting, which will arrest the biological process, thus giving physically stable but biologically unstable composts (de Bertoldi et al., 1983). On the other hand, high moisture value will prevent the ongoing composting activities (Tiquia et al., 1996). If in large composting system, high moisture content impacts on the system in two ways, it limit oxygen diffusion into the composting matrix and increases pliability of the material leading to matrix (Liang et al., 2003). This can result in increased aeration cost, reduced air permeability and may introduce anaerobic conditions from water logging in pore spaces impeding the composting processes.

Generally, optimum moisture contents for composting were range from 25% to 80% on a wet basis (w.b), with typical recommended values in the 50% to 70% range (Ahn et al., 2008). This relatively wide range of reported values shows that there is no universally applicable optimum moisture content for composting materials.

## 2.3 Oxygen concentration

Composting is an aerobic process, so an adequate supply of oxygen is a major consideration in controlling the composting process. Good control of oxygen concentration contributes to the results of the composting process, by affecting the, quality of end-product, energy consumption, and hygienization of compost. Oxygen concentration proved in the process has an interesting relationship to the rate of biodegradation. The oxygen concentration relates to (1) the main microbial groups during composting; (2) the balance of material; (3) the gas present in the internal atmosphere ($O_2$, $NH_3$, $CO_2$, $H_2S$); (4) phytotoxicity; and (5) pathogenic and indicator micro-organisms (de Bertoldi et al., 1988).

Oxygen exists within the compost matrix in both gaseous and dissolved form. But, only the dissolved form is available for use by microorganisms, since liquids are required as a medium for transporting oxygen in and out of the cell (Bongochgetsakul and Ishida, 2008). Oxygen is utilised as a terminal electron acceptor in micro organisms, for aerobic respiration and also in the oxidation of several sorts of organic substances in the mass (de Bertoldi et al., 1983). As permissible moisture content and oxygen availability are closely interrelated, Liang's (2003) study proved that moisture content was an important factor impacting on microbial activity by measuring the oxygen uptake during the composting process. Meanwhile, Richard's (2006) study proved the relationship between temperature and oxygen concentration. The lower temperature (<55 °C) rate gives highest $O_2$ concentration near ambient (21%) and dropped quickly between 1% and 4% $O_2$ (v/v). At 65 °C and occasionally 55 °C, the typical saturation relationship was frequently observed, to have higher degradation rates at lower $O_2$ concentrations. When the oxygen level falls below the ranges, anaerobic microorganisms begin to exceed aerobic, this makes fermentation and anaerobic respiration process take over. The intermediate fermentation products maybe accumulating at low of $O_2$ concentration under certain substrate and temperature condition, and priming the system for rapid $CO_2$ production (Diaz et al., 2007b).

Schulze (1962) also discussed oxygen consumption which is related to temperature in the composting process, by following relationship:

$$y = a10^{KT} \qquad [2.1]$$

where:
y = oxygen consumption rate, $mgO_2/g(VS).h$
a = 0.1, intercept at T = 0 °C
K = 0.028
T = temperature, °C

In reality y will not increase exponentially beyond a certain optimum temperature, and only valid in the temperature range of 20 to 70 °C. Schulze found that qualitatively, the closer the environmental condition approached an optimum level, the greater as the rate and oxygen uptake (Diaz et al., 2007b). In composting, dropped oxygen level can be improved by oxygen supplied by ventilation. The ventilation process is a part of the aeration system, to supply sufficient oxygen for aerobic respiration while performing associated heat removal. Ventilation, besides providing oxygen to the mass, also serves other function such as moisture and temperature control.

## 2.4    Structure

Rapid composting is dependent on condition and structure of the starting material. Size reduction in composting is necessary to increase the surface area of the material. The smaller the particle, the greater the ratio of the surface area to mass (de Bertoldi et al., 1983). This explained that the smaller particle size makes a better of biological degradation. The preparation of organic fractions for composting, of which there are two, mechanical or biological-mechanical, is an important start to the composting process. .After this process, the biodegradable organic fraction has been drastically conditioned and disintegrated, making it more easily separated by mechanical means from the inert material (de Bertoldi et al., 1983).

## 2.5    Microbial community

Composting is a self-heating process, an aerobic solid phase biodegradative process of organic waste material. In this composting process, microbial involved several interrelated factors, i.e metabolic heat generation, ventilation (oxygen output), temperature, moisture content, and available nutrient (de Bertoldi et al., 1983).

Generally, composting proceeds through three phase: 1) mesophilic (moderate temperature phase), 2) thermophilic (high temperature phase) and 3) cooling and maturation phase. During initial decomposition, a mesophilic microorganism is rapidly broken down by the soluble, into degradable compounds. Heat will produce and causes the compost temperature to rapidly rise, into the thermophilic phase. Recently, large varieties of mesophilic, thermotolerant, and thermophilic aerobic microorganisms, including bacteria, actinomycetes, yeasts and various other fungi have been extensively reported in composts (Faure and Deschamps, 1990, Chen et al., 2006, Déportes, 1998, Hassen et al., 2001, Suárez-Estrella et al., 2007).

Bacteria are more important than fungi from the beginning of process especially for the composting of sewage sludge. Bacteria are defined as fast composters and responsible for most of the decomposition. It stabilized and digests available nutrient as a product of decompositions. Mentioned by Hassen et al., (2001) mesophilic bacteria were the dominant degraders at an early phase of composting process with temperature of 20 – 40 °C.  These mesophilic bacteria were partially killed during the thermogenic stage, where on the contrary the thermophilic bacteria take their place this process (Nakasaki, 1985).

Fungi also play important role in the degradation of biomass and an intricate role in the early stage of composting. Reports that a modest variety of fungi was present over the course of the first 96h of composting (Hansgate et al., 2005). At this stage, fungi compete with bacteria for the easily available substrates. After the bacteria exceed the maximum specific growth rates, fungi are very soon out-competed (Bertoldi, 2007). However, they become abundant at the last stage of composting or in maturation process. As fungi benefit from the decrease in temperature, pH and moisture content that take place as the process evolves (de Bertoldi et al., 1983)

2.**6**    **Temperature**

Compost temperature is an important factor due to its influence on the activity of microorganisms such as microbial metabolic rate and population structure. Composting begins at the ambient temperatures and with a microbial community resident in the original organic material. High microbial activities will leads to a rapid increase in temperature and a period of elevated temperatures. During this thermophilic stage, many of the non-thermo tolerant organisms are activated, including several pathogens. This will be followed by a gradually decrease in microbial activity, and a cooling and maturation of the composting mass (Steger et al., 2007).

According to Stentiford (1996), temperatures above 55 °C are important to maximise sanitisation. Meanwhile, temperature between 45 and 55 °C are to improve the degradation rate and between 35 and 40 °C to increase microbial diversity. The optimum temperature for composting process is considered to be approximately 60 °C according to maximizing respiration rate and $CO_2$ evolution rate. Furthermore, in thermophilic range, the optimum temperature was determined to be 54 ºC by studying the effect on oxygen uptake rate, specific growth rate and enzymatic activity of microorganisms (Miyatake and Iwabuchi, 2006).

However, in some studies, composting also can be carried out in mesophilic temperatures lower than 45 °C (Tang et al., 2007). According to Liang et al., (2003), the maximums cumulative uptakes of $O_2$ was noted at a temperature of 43°C in different composting processes in the temperature range of 22–57°C. These indicated that mesophilic composting at lower temperature also favourable for the decomposition of waste (Tang et al., 2007). Mesophilic composting is described as more effective for the reduction of composting mass because of the higher decomposition activity of microbial activity. However, higher temperature is still considerable more effectively at elimination of pathogenic contaminants during composting (Hassen et al., 2001).

3    **Pathogens in composting**

The pathogen content in compost is important because, if improperly treated compost can be a source of pathogen to the environment and, as such, a threat to humans and animals. Reports in municipal wastes (Déportes, 1998, Hassen et al., 2001), sewage sludges (D J Dudley, 1980), and other organic sludge (Bustamante et al., 2008) shows they contain a different kinds of harmful pathogens. Sewage sludge is generally richer in pathogens than municipal solid waste (de Bertoldi et al., 1983). Pathogens found in composting can be viruses, bacteria, protozoa or helminths especially in municipal solids waste and sewage sludge. Pathogens are heat-sensitive; the heat increase occurs in the composting process and eliminates them, leading to a pathogen-free end product. This is also defined as disinfection or sanitization process (Déportes, 1998).

### 3.1 Pathogen control and inactivation

*Salmonella* spp. and *E. coli* are well known as pathogen indicators, in standard quality of composting. The role of both Salmonella sp. and *E. coli* in defining the ecological quality composts products was establish in the European Commission Decisions. The limits for their densities ($<10^3$ MPN g$^{-1}$ and Absence in 50 g, respectively) is fixed in order to assign a seal quality in Commission Decision 2001/688/EC;Commission Decision 2005/384/EC (Briancesco, 2008). Meanwhile, the UK composting standard (BSI, 2005) requires composts that are sold to be free of *Salmonella* spp. and to contain fewer than 1000 colony forming units of *E. coli* per gram of material. Summarized by Böhm et al., (2007) the selected species to be an indicator organisms should fulfil several requirement:

a. present with high probability in the raw material involved
b. the transmission via products must be a factor in epidemiology
c. if a biotechnology process is used the indicator should not be involved in the process itself
d. the indicator should not be an organisms that is generally present in soil and soil-related material
e. the method for isolation and identification must be simple definitive and reliable if applied to a substrate with a complex microbiological matrix such as compost sludge or related materials.

Among the bacterial indicators of faecal pollution, the role of *E. coli* must be considered because this bacterium has a good correlation with the presences of other enteric bacteria with the same characteristics. The Enterococci which belongs to the same group as *E. coli,* is known to have a higher resistance to physico-chemical treatments and has a close connection to the final gut tract of warm-blooded animals, which could be valid faecal indicator group. The detection of *Salmonella* spp, is very useful in stabilization of compost material. It is related to human pathogen and very important in order to evaluate the sanitary quality of stable compost to limit the health risk (Briancesco, 2008).

The destruction of pathogen during the composting period can be effected by a number of mechanisms such as exposure to heat, temperature/time exposure, competition for nutrients, microbial antagonism (including antibiotic production and direct parasitism), production of organic acids and ammonia (Pereira-Neto et al., 1986, Wilkinson, 2008). Wilkinson, (2008) described the function of temperature and the length of exposure as the most important in the inactivation of pathogen. Besides, the temperature is relatively easy to measure during composting.

Exposure to an average temperature during composting of about 55-60°C for at least 3 days is usually sufficient to kill the vast majority of enteric pathogen (Deportes et al., 1995). Table 3.1 below shows the destruction of the pathogens depends on the temperature reached and the duration of the oxidation process.

Table 3.1: Level of temperature and length of time necessary to destroy some pathogen present in primary products of composts (Deportes et al., 1995)

| Organisms | Lethal temperature and necessary time |
|---|---|
| *Salmonella* spp | 15-20 at 60 $^{o}$C; 1 h at 55 $^{o}$C |
| *Escherichia coli* | 15-20 at 60 $^{o}$C; 1 h at 55 $^{o}$C |
| *Entamoeba hystolitica* | 68 $^{o}$C; time not given |
| *Taenia saginata* | 5 min at 71 $^{o}$C |
| *Necator americanus* | 50 min at 45 $^{o}$C |
| *Shigella* spp | 1 h at 55 $^{o}$C |

Note: the typical range of composting is 55-65 $^{o}$C

## 3.2    Regrowth of pathogen in compost

The enteric pathogen is known to be able to grow after having diminished below the detectable limits and representing a health hazard for certain uses of compost. Regrowth of pathogens in compost depends upon a number of factors such as moisture content bio-available nutrients, temperature and indigenous micro organisms (Sidhu et al., 2001). The most critical factor to prevent regrowth of pathogen is stability. The compost materials appear to be stable when it is too dry to support high rates of microbiology activity. Re-wetting of these compost material will lead an ideal environment being provided for pathogens to repopulate (Wilkinson, 2008).

A suggestion by Böhm et al., (2007) is that the compost product should be stored under dry conditions, as most bacteria will not propagate in a dry product even if the material seems to be stable and dry. Besides, the vapour pressure within the product should be in balance with relative humidity of around 90%. Long-time storage (for example 2 years) of composted biosolids to achieve compost stability and reduce pathogen regrowth potential, cannot guarantee the bio-safety of composted biosolids. Thus, it became more likely to promote *Salmonella* regrowth, as a result of the declining effect of indigenous antagonistic micro-organisms (Sidhu et al., 2001).

*Salmonella* serovar Enteritidis was observed to survive for at least 3 months in the stabilized compost after 8 weeks of composting. Reports show that industrial biowastes composts are usually commercialized as product after 4 to 6 months (Lemunier, 2005). Thus, the long term duration storage for composting should be avoided and the other tests with older industrial composts should be done to confirm the potential risk such as survival and regrowth of pathogen. Nevertheless, the proper management of both of the thermophillic and maturation phase during the composting process is still a critical step in controlling the pathogen survival.

4      Conclusion

Composting is a technique that can be used to reduce the environmental problems related to waste disposal. A key factor in the process is the reduction of pathogenic micro organisms Thus, the relation between process of composting and pathogen is tremendously important. In order to reduce pathogenic micro organisms to an acceptable level it is important for the process to be well managed and controlled in all the production steps.